# SCALABLE, ROBUST AND REAL TIME COMMUNICATION ARCHITECTURE FOR WIRELESS SENSOR NETWORKS

[1]Deepali Virmani and [2]Satbir Jain
[1]Department of CSE GPMCE, Delhi.
[2]Department of Computer Science of Engineering NSIT, Delhi.
INDIA
deepalivirmani@gmail..com

**ABSTRACT**

*In this paper we propose wireless sensor network architecture with layered protocols, targeting different aspects of the awareness requirements in wireless sensor networks. Under such a unified framework, we pay special attention to the most important awareness issues in wireless sensor networks: namely dynamic awareness, energy awareness and spatiotemporal awareness. First, we propose the spatiotemporal aware routing protocol for wireless sensor networks, which maintains a constant delivery speed for soft real-time communication. Second, we introduce the time-energy aware aggregation scheme, which increases the degree of aggregation and reduces energy consumption without jeopardizing the end-to-end delay. It is the data aggregation scheme to take the timely delivery of messages as well as protocol overhead into account to adjust aggregation strategies adaptively in accordance with assessed traffic conditions and expected wireless sensor network requirements. Third, we also deal with dynamic awareness by proposing a new communication category based on a lazy-binding concept. The evaluation of this integrated architecture demonstrates its performance composability and its capability to be tailored for applications with different awareness requirements. We believe that the integrated architecture, supporting various kinds of awareness requirements, lays a foundation for overall wireless sensor network.*

## 1) INTRODUCTION

A wireless sensor network is a collection of nodes organized into a cooperative network [2]. Each node consists of processing capability (one or more microcontrollers, CPUs or DSP chips), may contain multiple types of memory (program, data and flash memories), have a RF transceiver (usually with a single omni-directional antenna), have a power source (e.g., batteries and solar cells), and accommodate various sensors and actuators. The nodes communicate wirelessly and often self-organize after being deployed in an ad hoc fashion. Systems of 1000s or even 10,000 nodes are anticipated. These tiny sensor nodes, which consist of sensing, data processing, and communicating components, leverage the idea of sensor networks based on collaborative effort of a large number of nodes. A wireless sensor network (WSN) is a computer network consisting of many, spatially distributed devices using sensors to monitor conditions at different locations, such as temperature, sound, vibration, pressure, motion or pollutants[1][2][5]. The development of wireless sensor networks was originally motivated by military applications such as battlefield surveillance. However, wireless sensor networks are now used in many civilian application areas, including environment and habitat monitoring, healthcare applications, home automation, and traffic control[1][2][3]. Usually these devices are small and inexpensive, so that they can be produced and deployed in large numbers. The size and price requirements imply that the devices resources in terms of energy, memory, computational speed and bandwidth are severely constrained. The concept of wireless sensor networks is based on a simple equation: Sensing + CPU + Radio = Thousands of potential applications wireless Sensor networks involve three areas: sensing, communications, and computation (hardware, software, algorithms). Very useful technologies are wireless database technology such as queries, used in a wireless sensor network, and network technology to communicate with other sensors, especially multihop routing protocols.The emerging field of wireless sensor networks combines sensing, computation, and communication into a single tiny device. Through advanced mesh networking protocols, these devices form a sea of connectivity that extends the reach of cyberspace out into the physical world. The mesh networking connectivity will seek out and exploit any possible communication path by hopping data from node to node in search of its destination. While the capabilities of any single device are minimal, the composition of hundreds of devices offers radical new technological possibilities. The power of wireless sensor networks lies in the ability to deploy large numbers of tiny nodes that assemble and configure themselves. Usage scenarios for these devices range from real-time tracking, to monitoring of environmental conditions, to ubiquitous computing environments, to monitoring of the health of structures or equipment. While often referred to as wireless sensor networks, they can also control actuators that extend control from cyberspace into the physical world.[2] The most straight forward application of wireless sensor network technology is to monitor remote environments for low frequency data trends. For example, an office could be easily monitored for fire in any of its room by hundreds of sensors that automatically form a wireless interconnection network and immediately report the detection of any fire. Unlike traditional wired systems, deployment costs would be minimal. Instead of having to deploy thousands of feet of wire routed through protective installers simply have to place quarter-sized device at each sensing point. The network could be incrementally extended by simply adding more devices no rework or complex configuration. In addition to drastically reducing the installation costs, wireless sensor networks have the ability to dynamically adapt to changing environments. Adaptation mechanisms can respond to changes in network topologies or can cause the network to shift between drastically different modes of operation. For example, the same embedded network performing leak monitoring in a chemical factory might be reconfigured into a network designed to localize the source of a leak and track the diffusion of poisonous gases. The network could then direct workers to the safest path for emergency evacuation. Current wireless systems only scratch the surface of possibilities emerging from the integration of low-power communication, sensing, energy storage, and computation. Unlike traditional wireless devices, wireless sensor nodes do not need to communicate directly with the nearest high-power control tower or base station, but only with their local peers. Instead, of relying on a pre-deployed infrastructure, each individual sensor or actuator becomes part of the overall infrastructure. Peer-to-peer networking protocols provide a mesh-like interconnect to shuttle data between the thousands of tiny embedded devices in a multi-hop fashion. As soon as people understand the capabilities of a wireless sensor network, hundreds of applications spring to mind. It seems like a straightforward combination of modern technology. However, actually combining sensors, radios, and CPU's into an effective wireless sensor network requires a detailed understanding of the both capabilities and limitations of each of the underlying hardware components, as well as a detailed understanding of modern networking technologies and distributed systems theory. Each individual node must be designed to provide the set of primitives necessary to synthesize the interconnected web that will emerge as they are deployed, while meeting strict requirements of size, cost and power consumption. A core challenge is to map the overall system requirements down to individual device capabilities, requirements and actions. To make the wireless sensor network vision a reality, architecture must be developed that synthesizes the envisioned applications out of the underlying hardware capabilities.

## 2) THE NEED FOR NEW SENSOR NETWORK ARCHITECTURE

Wireless Sensor Networks (WSN) has emerged as a new information-gathering paradigm based on the collaborative efforts of a large number of self-organized sensing nodes. These networks form the basis for many types of smart environments such as smart hospitals, intelligent battlefields, earthquake response systems, and learning environments. A set of applications, such as biomedicine, hazardous environment exploration, environmental monitoring, military tracking and reconnaissance surveillance, are the key motivations for the recent research efforts in this area. In sensor networks, nodes are deployed into an infrastructure free environment. Without any a priori information about the network topology or the global, even local view, sensor nodes must self-configure and gradually establish the network infrastructure from the scratch during the initialization phase. With the support of this infrastructure, nodes are able to accept queries from remote sites, interact with the physical environment, actuate in response to the sensor readings, and relay sensed information through the multi-hop sensor networks. Different from traditional networks, sensor networks do impose a set of new limitations for the protocols designed for this type of networks. Devices in sensor networks have a much smaller memory, constrained energy supply, less process and communication bandwidth.[2][4][8] Topologies of the sensor networks are constantly changing due to a high node failure rate, occasional shutdown and abrupt communication interferences. Due to the nature of the applications supported, sensor networks need to be densely deployed and have anywhere from thousands to millions of sensing devices, which are the orders of magnitude larger than traditional ad hoc mobile networks. In addition, energy conservation becomes the center of focus due to the limited battery capacity and the impossibility of recharge in the hostile environment. With such a vast difference between traditional networks and sensor networks, it is not appropriate and inefficient to port previous solutions for ad hoc networks into sensor networks with only incremental modifications. For instance, the sheer number of sensor nodes makes flooding-based standard routing schemes (e.g. DSR [9]and AODV[10] for ad hoc networks undesirable. Although applications for sensor networks remain diverse, one commonality they all share is the need for a network infrastructure tailored for sensor networks. Without a scalable routing service, broadcast storms caused by the route discovery may result in significant power consumption and possibly a network meltdown. Without a real-time communication service, applications cannot react to the changes of the environment quickly enough to be effective. Without efficient energy-aware design, nodes in the sensor networks could deplete themselves after only several rounds of burst activities. Without fault-tolerance and self-stabilization supports in such a dynamic and faulty system, sensor networks could never converge and are unable to guarantee an effective transport service to the applications. We do not close the eyes to the importance of the killer applications, but we argue that without a new network architecture tailored to the characteristics of sensor networks, the popularity of sensor networks cannot be a reality in the near future.

2.1) Proposed Features of New Sensor Network Architecture

As mentioned in previous sections, the unique features of sensor networks necessitate a new set of novel solutions tailored for this type of system. In this section, we specifically identify four essential topics for the sensor architecture . On the one hand, these topics are orthogonal because they reside in the different layers of the network architecture we propose and the solutions of those topics can work independently with each other. On the other hand, these topics are associated in the sense that they are complementary to each other in order to provide an integrated solution for sensor networks.

2.1.1) Soft Real-time Communication for Timely Response

To date, few results exist for sensor networks that adequately address real-time requirements for time-critical applications, such as battlefields and earthquake response systems. The correctness of such systems not only depends on the logical correctness of the results, but also the time when such results are produced. Without real time guarantee, actions taken by these systems would not be as effective as they should be. For example, in a target tracking system, an intruder's location should be reported to pursuers within 10 seconds bound so that pursuers can take effective actions. Since sensor networks deal with the real world and communication delays within sensing and actuating loops directly affect the response time of the applications, it is often necessary for communication to meet real-time constraints.

We identify our first feature for new sensor network architecture  as providing a real-time communication service that can support soft real-time end-to-end delivery under unpredictable network environments. Specially, following features  should be addressed:
• A novel mechanism to provide probabilistic soft real-time end-to-end delay guarantee, while underlying MAC only supports the contention-based best effort packet forwarding.
• A mechanism to estimate accurately the transient and long-term utilizations of wireless networks for the purpose of packet admission, scheduling and forwarding
• An integrated solution to reduce network congestion through multiple feedback-based adaptations
• An enhanced differentiated scheme to meet real-time constraints for packets with different priorities
• A routing scheme that is pure decentralized in order to cope with scalability issues, while at the same time avoiding system wide race condition and instability
The proposed feature will utilize a feedback-based congestion control to guarantee packet delivery speed across the network. With such a support, applications can estimate an end-to-end delay before making admission decisions and dynamically adjust the workload they generate to meet their real-time requirements.

2.1.2) Data Aggregation for Congestion Control and Energy Conservation

Data aggregation techniques are proposed to address multiple issues. Data aggregation performed among a group of nodes can effectively reduce total amount of application data shipped out, thus reduce network congestion and energy consumption. To the best of our knowledge, all recent research focuses on application dependent data aggregation techniques (ADDA), in which aggregation heavily depends on the application layer information. By placing a naming (semantic) restriction on the aggregated data, those techniques impose that lower layer protocols must have knowledge of these naming semantics and limit the types of data that can be aggregated. For example, the aggregation module must have the knowledge that the temperature readings from the northeast corner of a network should not be combined with the temperature from the southwest corner just because they share a common type and make sure aggregation is not performed between the messages containing temperature readings and messages containing acoustic readings. In view of those limitations, we focus on an application independent data aggregation (AIDA) approach. This approach isolates aggregation decisions from application specifics by performing adaptive aggregation in an intermediate layer that resides between the traditional data-link and network layer. In order to reduce network congestion and achieve a high degree of energy conservation, we should address following features
• A modular architecture to isolate aggregation decisions effectively from application specifics
• An approximate model for the MAC contention for the purpose of control
• An adaptive aggregations scheme based on the feedback on network situations
• A novel mechanism to increase the degree of aggregation without jeopardizing the end-to-end delay
• An enhance scheme to incorporate real-time guarantees and differentiated QoS supports into this aggregation framework

Our feature is expected to improve the efficiency in bandwidth utilization, a resource that is most precious in sensor networks. The auxiliary benefit of our solution is the energy-conservation by reducing packet collisions and control overhead.

2.1.3) Robust Data Delivery under Failure and Mobility
Sensor networks are faulty networks where failures should be treated as normal phenomena. Unreliable nodes, constrained energy, high channel bit error ratio, interference and jamming, multi-path-fading, asymmetric channel and weak security make the communication highly unreliable. At same time, sensor networks are highly dynamic networks where network topologies are constantly changing due to a high rate of node failure, changes of power modes, and nodes' mobility. It is a challenge research problem to provide a robust data delivery under such a situation. Previous protocols proposed need to update and maintain routing tables or at least neighborhood tables for the purpose of routing, and they suffer delay to establish and maintain these tables if the network is highly dynamic. With constant node failures and frequent message loss, the state-sensitive routing protocols such as AODV, DSR and DSDV take a huge amount of time and energy to stabilize. Acknowledging that state-based solutions are inefficient to cope with highly dynamical sensor networks, we proposed a feature that is altogether state-free for robust data delivery. In this solution, we aim at providing not only a reliable communication scheme, but also a fast response and recovery from the failures with a much less control overhead. Specifically following issues will be addressed in our architecture

• A swift & self-stabilizing approach to deal with instability caused by fast flow dynamics inside networks such as nodes' failure and mobility

• An efficient approach to reduce the inconsistency between outdated routing information a node keeps and the volatile network situations with minimal overhead

• A reliable scheme which prevents the performance degradations in packet delivery, end-to-end delay and control overhead, while allowing nodes going to a dormant state in order to conserve energy effectively

With robust data delivery support, On the one hand, applications can deliver data more reliable and fast in face of high node failure rate and mobility. On the other, less bandwidth and energy will be consumed by our novel state-free solution.

2.3.4) Localization for Routing Scalability

When a system scales up to thousands and even millions of nodes, a set of pervious solutions is no longer applicable. For example, DSR [9] and AODV [10] are the standard routing schemes for ad hoc wireless networks with up to hundred nodes. Those protocols depend on a so-called on-demand routing discovery with flooding to find an end-to-end path to a destination. The sheer number of sensor nodes makes such a global flooding undesirable. When thousands of nodes communicate with each other, broadcast storms may result in significant power consumption and possibly a network meltdown. Location awareness is an essential building block for sensor networks. Besides tagging the sensor data with location context, localization is commonly used for routing scalability. Location-based routings are widely accepted as de fact standard routing technique for sensor networks. This is because that the location-based routing schemes only need to keep neighbour information regardless the scale of the networks. Moreover, the location-address communication precludes the requirement of the route discovery, otherwise needed by ad hoc network protocols. In order to achieve scalability in routing, this architecture addresses the localization problem through which nodes can find out their location and perform location-based routing. Roughly, localization techniques in sensor networks can be divided into to two major categories: range-based localization and range-free localization. The former is defined by protocols that use absolute point-to-point distance or angle estimates for estimating location. The later make no assumption about the availability or validity of such information. Range-based localizations are widely investigated in recent years. Such technique yells better precision under control environment or by using sophisticated devices. Much less research has be done on range-free localization, which are regard as an cost-effective and sufficient solution for sensor networks without costly hardware requirements. Our range-free localization work will address following features

• A novel cost-effective localization scheme without sophisticated hardware requirements

• An evaluation on the impacts of localization error to the performance of sensor network applications

• A theoretical model to estimate localization error bound and identify optimal system configuration

Our work in this topic will investigate these issues and design a scalable localization algorithm with enhanced performance over pervious solutions

3) PROPOSED NEW SENSOR NETWORK ARCHITECTURE

Physical Layer with Power Management: This energy-aware layer is responsible for node duty-cycle scheduling for power conservation and sensing coverage management, which provide full sensing coverage to a geographic area while at the same time minimize energy consumption and extend system lifetime by leveraging the redundant deployment of sensor nodes.

Robust & self MAC Layer: This layer solves the issues related to robustness and self stabilization. It can provide per-hop reliability if required by higher layer. Moreover, this layer deals with node failure and mobility issues internally as much as possible before signaling the network layer for further assistances.

Application Independent Data Aggregation Layer: This time-energy aware layer provides data aggregation in order to reduce control overhead in MAC layer and energy consumption in radio communication. Adaptive control is provided in order to achieve aggregation without jeopardizing the end-to-end delay.

Differentiated Packet Scheduling Layer: This QoS-time-aware layer supports differentiated forwarding service among the packets with different priorities. The criterion for the differentiation includes not only the time constraints but also the spatial constraints.

Soft Real-Time Routing Layer: This time-aware scalable routing layer provides soft real-time communication for end-to-end packet delivery. Network congestion control is achieved by localized rerouting and traffic policing to the lower layer. Constant delivery speed is maintained through a combination of non-deterministic forwarding and neighborhood-based feedback control. The scalability issue is solved by the location-based routing and localized control scheme.

Transport layer: This layer abstracts communication endpoints into entities. This layer maintains robust connections between entities while both ends are mobile.

Localization Service: Localization is a cross layer service for this network architecture as a whole. It provides location information for 1) the sensing coverage management, 2) the velocity calculation for the differentiated packet scheduling, 3) the location-address soft real-time routing, 4) the entity formation and 5) the location service for sensor network applications such as the enemy tracking and temperature mapping.

## 4) CONCLUSIONS &FUTURE WORK

Recent advances in the development of the low-cost, power-efficient embedded devices, coupled with the rising need for support of new information processing paradigms such as smart spaces and military surveillance systems, have led to active research in large-scale, highly distributed sensor networks of small, wireless, low power, unattended sensors and actuators. A set of applications, such as biomedicine, hazardous environment exploration, environmental monitoring, military tracking and reconnaissance surveillance, becomes the key motivations of the recent rise in research efforts in this area. While applications keep diversifying, one common property they share is the need for an efficient network architecture tailored toward sensor networks. Previous solutions designed for traditional networks serve as good references; however, due to the vast differences between previous paradigms and sensor networks, the direct use of these more classical solutions lead to inefficiencies, overkill, or non-functioning systems. This proposal aims at **building scalable, robust and time energy aware network architecture for sensor networks**. The topics we address cover every layer of the network architecture, paying special attention to stateless real-time routing, application independent data aggregation, robust data delivery and range-free localization. The integrated network architecture based on these solutions lays a foundation for overall sensor network research and helps bring this new paradigm into reality.

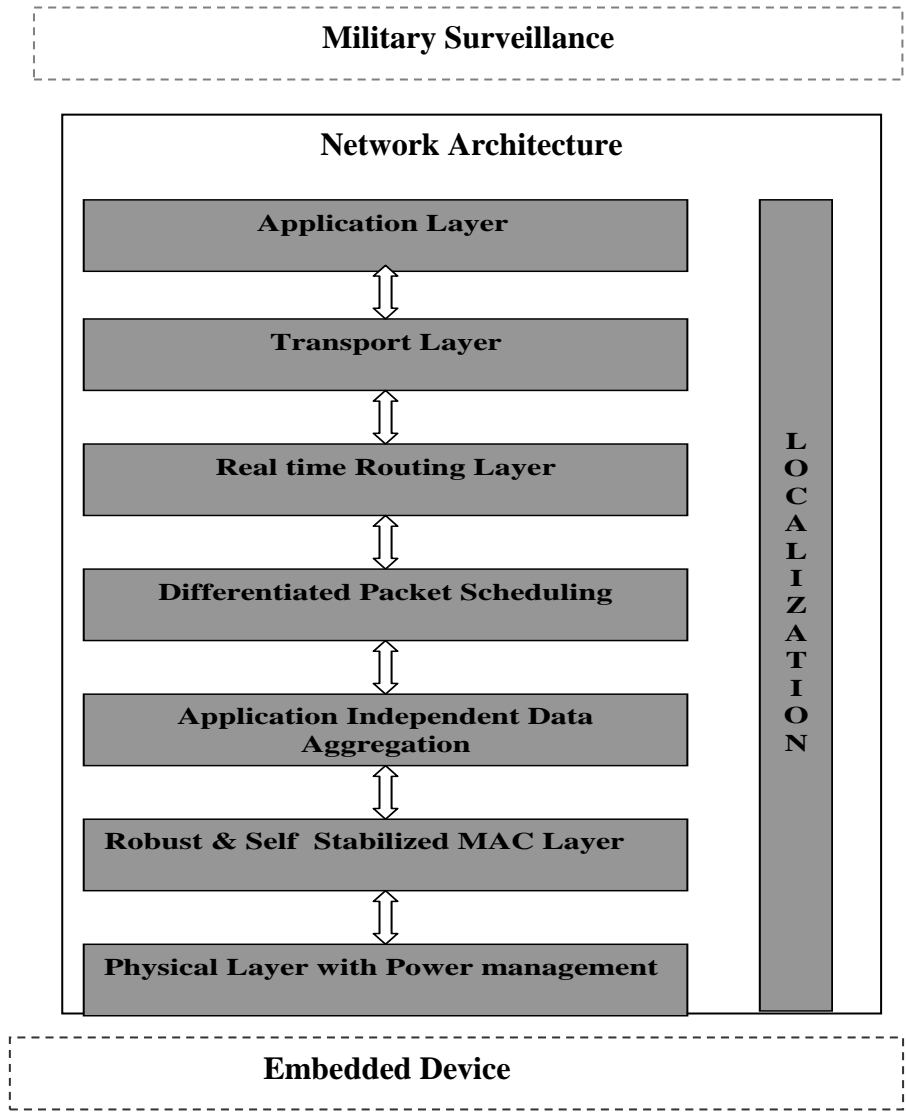

Fig 1: Proposed New Network Architecture